\documentstyle[12pt]{article}
\if@twoside  m
\else
\fi
\newcommand{\ie}{{\em i.e.}}
\newcommand{\eg}{{\em e.g.}}
\newcommand{\QED}{\mbox{\rule[-1.5pt]{6pt}{10pt}}}
\newcommand{\lhs}{{\em lhs }}
\newcommand{\rhs}{{\em rhs }}
\newcommand{\re}{{\rm Re\,}}
\newcommand{\im}{{\rm Im\,}}
\newcommand{\card}{{\rm card\,}}
\newcommand{\C}{C\!\!\!\rule[.5pt]{.7pt}{6.5pt}\:\:}
\newcommand{\R}{I\!\!R}
\newcommand{\Z}{Z\!\!\!Z}
\newcommand{\BB}{{\cal B}}
\newcommand{\DD}{{\cal D}}
\newcommand{\EE}{{\cal E}}
\newcommand{\HH}{{\cal H}}
\newcommand{\II}{{\cal I}}
\newcommand{\KK}{{\cal K}}
\newcommand{\LL}{{\cal L}}
\newcommand{\NN}{{\cal N}}
\newcommand{\RR}{{\cal R}}
\newcommand{\VV}{{\cal V}}
\newcommand{\XX}{{\cal X}}
\begin{document}
\title{Magnetoresonances on a lasso graph}
\date{}
\author{Pavel Exner}
\date{\normalsize Nuclear Physics Institute, Academy of Sciences, \\
25068 \v Re\v z near Prague, and \\ Doppler Institute, Czech Technical
University, B\v rehov{\'a} 7, \\ 11519 Prague, Czech Republic }
\maketitle

\begin{quote}
We consider a charged spinless quantum particle confined to a
graph consisting of a loop to which a halfline lead is attached; this
system is placed into a homogeneous magnetic field perpendicular to
the loop plane. We derive the reflection amplitude and show that
there is an infinite ladder of resonances; analyzing the resonance
pole trajectories we show that half of them turn into true embedded
eigenvalues provided the flux through the loop is an integer or
halfinteger multiple of the flux unit $\,hc/e\,$. We also describe a
general method to solve the scattering problem on graphs of which the
present model is a simple particular case. Finally, we discuss ways
in which a state localized initially at the loop decays.
\end{quote}
\vspace{10mm}

\noindent
I begun my career at times when the world was much less connected,
and of most parts we knew only from journals arriving not quite
regularly. My first encounter with the mathematical theory of
unstable quantum systems and related scattering problems occurred in
this way, particularly through \cite{HLM,HM1} and related papers by
Larry Horwitz; the subject remained for me as well as for him an old
love to which we return regularly from time to time. Only many years
later I had an opportunity to met him in person and to appreciate
also his charisma.  A distinguished birthday is a good opportunity to
come with another decay scattering system; I should say that I prefer
presents which are amusing rather than expensive.

\section{Introduction}

Quantum mechanics for a nonrelativistic particle whose configuration
space is a graph has been studied already a long time ago in
connection with models of organic molecules \cite{RuS}. A recent new
interest to these problems \cite{Ad,AL,AEL,ARZ,E2,E3,EG,ES,GP,GLRT}
has been stimulated, in particular, by the progress of experimental
solid state physics which allows us to produce semiconductor
``quantum wire" structures and other ``mesoscopic devices"; quantum
mechanical graphs represent a natural model for many of them.

Graph systems provide a convenient mean to study various quantum
effects both from the theoretical and experimental points of view,
because the freedom in setting the geometry of the configuration
space allows us to create different dynamics; at the same time,
models of this type are often explicitly solvable. This concerns, in
particular, resonance scattering effects associated with the existence
of quasi\-stationary states in graph loops and appendices --- see, \eg,
\cite{ESe}. These effects fit well, of course, into the general
theory of decay scattering systems as exposed in \cite[Chaps.
1,3]{E1}, \cite{HS}, or \cite[Sec.XII.6]{RS}, but they also make it
possible to illustrate it and to draw fully specific conclusions.

Our aim here is to investigate one more solvable model of this type.
It consists of a halfline attached to a loop placed into a magnetic
field; the parameters are the magnetic flux through the loop and
``coupling strengths" between the graph links at the junction. Our
analysis differs from an earlier treatment of similar systems
\cite{Bu,JS} in several aspects. First of all, we consider a
different and more general coupling between the loop and the
halfline, and we put emphasis on the analytical solution of the
problem. Furthermore, we shall be concerned with the decay and
scattering properties of the system rather than with persistent
currents induced by the magnetic field.

Let us review briefly the contents of the paper. The Hamiltonian of
the model we are going to study is introduced in the following
section. Next, we derive in Section~3 its spectral and scattering
properties. Then we make a digression and describe a general method
to treat scattering problem on an arbitrary graph by ``discretizing"
it, \ie, transforming the corresponding Schr\"odinger equation into a
set of linear equations involving just the wavefunction values at the
graph nodes. Returning to our model, we analyze in Section~5 its
resonance structure by deriving an explicit expression for the
resolvent and finding the resonance--pole trajectories. Finally, in
the concluding section we treat our model as a decay system and show
how a state localized initially at the loop decays (or does not
decay) in the course of time.

\section{Description of the model}

   \begin{figure}
   % INSERT FIGURE 1
   \caption{A lasso graph}
   \end{figure}

Consider a quantum particle confined to the lasso--shaped graph
$\,\Gamma\,$ sketched on Figure~1, \ie, a circular loop of radius
$\,R\,$ to which a halfline lead is attached. We suppose that the
particle is nonrelativistic, spinless, and charged. To be specific we
assume that its charge is $\,q=-1\,$; we adopt the usual rational
system of units, $\,e=c=2m=\hbar=1\,$. The system is placed into a
homogeneous magnetic field of intensity $\,B\,$; the vector potential
$\,\vec A\,$ can be chosen tangent to the loop with the modulus
   \begin{equation} \label{vector potential}
A\,=\, {1\over 2}\,BR\,=\, {\Phi\over L}\,,
   \end{equation}
where $\,\Phi\,$ is the magnetic flux through the loop and $\,L\,$ is
the loop perimeter. Under the convention we have adopted the natural
flux unit \cite{Bu} is $\,hc/e= 2\pi\,$, so the \rhs of (\ref{vector
potential}) can be also written as $\,\phi/R\,$ where $\,\phi\,$ is
the flux value in this scale.

The state Hilbert space of the model is $\,\HH\equiv L^2(\Gamma):=
L^2(0,L)\oplus L^2(\R_+)\,$; the wave functions will be written
as columns, $\,\psi= {u \choose f}\,$. To construct the Hamiltonian
we begin with the operator describing the free motion on the loop and
the lead under the condition that the graph vertex is ``fully
disconnected", so $\,H_{\infty}= H_{\rm loop}(B)\oplus H_{\rm
halfline}\,$, where
   \begin{equation} \label{disconnected parts}
H_{\rm loop}(B)\,=\, (-i\partial_x+A)^2\,, \qquad H_{\rm
halfline}\,=\, -\partial^2_x
   \end{equation}
with the Dirichlet condition, $\,u(0)=u(L)=f(0)=0\,$; if there
is no danger of misunderstanding we abuse the notation and employ the
same symbol for the arc--length variable on both parts of the graph.
The operator $\,H_{\rm loop}\,$ has a simple discrete spectrum; the
eigenfunctions
   \begin{equation} \label{chi_n}
\chi_n(x)\,=\, {e^{-iAx}\over\sqrt{\pi R}}\, \sin\left( nx\over
2R\right)\,, \qquad n=1,2,\dots
   \end{equation}
correspond to the eigenvalues $\,\left(n\over 2R\right)^2$, which are
embedded into the continuous spectrum of $\,H_{\rm halfline}\,$
covering the interval $\,[0,\infty)\,$. Notice that the effect of the
magnetic field on the {\em disconnected} loop amounts to a unitary
equivalence,
   \begin{equation} \label{u equivalence}
H_{\rm loop}(B)\,=\, U_{-A} H_{\rm loop}(0) U_A\,,
   \end{equation}
where $\,(U_A u)(x):= e^{iAx} u(x)\,$.

To couple the graph parts one has to follow the standard strategy
\cite{ES} which means to replace Dirichlet by a ``connected"
boundary condition at the vertex. In general, there is a
nine--parameter family of such conditions. This is too many; we
will be concerned with its three--parameter subfamily \cite{ES,ESe},
in particular, with a one--parameter set of boundary conditions known
as $\,\delta$--coupling \cite{E3}. Hence the Hamiltonian of our model
acts as the free operator specified by (\ref{disconnected parts}),
   \begin{equation} \label{Hamiltonian}
H_{\alpha,\mu,\omega}(B) {u \choose f} = {-u''\!-2iAu'\!+A^2u \choose
-f''}\;;
   \end{equation}
the wave function is continuous on the loop,
   \begin{equation} \label{loop continuity}
u(0)\,=\, u(L)\,,
   \end{equation}
and satisfies the requirements
   \begin{eqnarray}
f(0) \!&=&\! \omega u(0)+ \mu f'(0)\,, \nonumber \\
\label{4 parameter bc} \\
u'(0)-u'(L) \!&=&\! \alpha f(0)- \omega f'(0)\,, \nonumber
   \end{eqnarray}
for an $\,\alpha,\mu\in\R\,$ and $\,\omega\in\C\;$; the values of
$\,u,\,f\,$ and their derivatives at the vertex are understood as the
appropriate one--sided limits. However, we shall restrict ourselves
to the case of time--reversal invariant couplings which means to
assume that $\,\omega\,$ is also real; it has the meaning of a
coupling constant between the loop (with a point interaction) and the
halfline. In physical terms the conditions (\ref{loop continuity})
and (\ref{4 parameter bc}) express the conservation of probability
flow at the junction.

The $\,\delta$--coupling corresponds to the choice $\,\mu=0\,$ and
$\,\omega=1\,$ in which case the wavefunction is fully continuous,
   \begin{equation} \label{H continuity}
u(0)\,=\, u(L)\,=\, f(0)\,,
   \end{equation}
and
   \begin{equation} \label{H conservation}
u'(0)-u'(L)+f'(0)\,=\, \alpha f(0)\;;
   \end{equation}
for the sake of simplicity we shall write $\,H_{\alpha}\equiv
H_{\alpha,0,1}\,$.  The parameter $\,\alpha\,$ is a coupling constant
between the {\em disconnected} loop and the halfline; the fully
decoupled case corresponds to $\,\alpha=\infty\,$ as the notation
suggests.
\vspace{2mm}

\noindent
{\bf Remarks.} (a) The choice of the coupling at the vertex
corresponds to a conceivable quantum--wire experiment. There is an
approximation result \cite{E4} which means that the
$\,\delta$--coupling constant $\,\alpha\,$ can be regarded as a mean
value of a sharply localized potential. This corresponds, \eg, to a
screened electrode placed at the vicinity of the junction; in a
similar way one can model some of the more general boundary
conditions (\ref{loop continuity}) and (\ref{4 parameter bc})
relating the parameters to physical quantities which an
experimentalist can tune.

(b) In general, the vector potential enters the boundary conditions
--- see \cite{ARZ} and the remarks in Section~4.3 below. In the
present case, however, the outward tangent components of $\,\vec A\,$
at the junction have opposite signs, so their contributions cancel.
This may not be true if the loop is noncircular and has corners or
cusps, but one can always achieve a cancellation by a suitable gauge
choice. If the loop is viewed from outside as in the scattering
process, the only quantity which matters is the magnetic flux
$\,\Phi\,$ threading it.

(c) The S--matrix for a coupling of three semiinfinite wires
equivalent to (\ref{loop continuity}) and (\ref{4 parameter bc}) was
derived in \cite{ES}. This comparison shows, in particular, that
choosing $\,\alpha=\mu=0\,$ and putting $\,\epsilon:= \left(
2\omega\over 2+\omega^2 \right)^2$, we obtain the coupling used in
\cite{Bu}. On the other hand, the authors of \cite{JS} worked with
the ideal $\,\delta$--coupling, $\,\alpha=0\,$.

\section{Scattering and bound states}

\setcounter{equation}{0}

Consider now the scattering problem on $\,\Gamma\,$, \ie, the
reflection of a particle traveling along the halfline from the
magnetic--loop end. We limit ourselves to the stationary formulation
looking for generalized eigenvectors, in other words, solutions of
the equation $\,H_{\alpha}(B)\psi= k^2\psi\,$ which satisfy the
definition domain requirements with exception of global square
integrability. In view of (\ref{Hamiltonian}), the most general
Ansatz for such a solution is
   \begin{equation} \label{Ansatz}
u(x)\,=\, \beta\, e^{-iAx} \sin(kx+\gamma)\,, \qquad f(x)\,=\,
e^{-ikx}\! +r\,e^{ikx}
   \end{equation}
with ($\,k$--dependent) parameters $\,r,\,\beta\,$, and $\,\gamma\,$;
the latter is generally complex.

To find them we employ the boundary conditions. The identity
(\ref{loop continuity}) in combination with (\ref{vector potential})
leads to the relation
   \begin{equation} \label{gamma}
\tan\gamma\,=\, {\sin kL\over e^{i\Phi}- \cos kL}\,.
   \end{equation}
The conditions (\ref{4 parameter bc}) yield then a system of two
linear equations for $\,r, \beta\,$ which is solved by
$$
r\,=\, -\,{(1+ik\mu)\left\lbrack \alpha\,-\, {\RR\over \sin\gamma}\,
\right\rbrack +i\omega^2k  \over (1-ik\mu)\left\lbrack \alpha\,-\,
{\RR\over \sin\gamma}\, \right\rbrack -i\omega^2k}
$$
with
$$
{\RR\over\sin\gamma}\,=\, k\,\cos\gamma -iA\,\sin\gamma -\, e^{-\Phi}
\left\lbrack k\,\cos(kL+\gamma)-iA\, \sin(kL+ \gamma) \right
\rbrack\,.
$$
Using again (\ref{loop continuity}) and (\ref{gamma}), we arrive after a
simple algebra at the expression
   \begin{equation} \label{reflection}
r(k)\,=\, -\,{(1+ik\mu)\left\lbrack \alpha\,-\, {2k\over \sin
kL}\,(\cos\Phi-\cos kL)\, \right\rbrack +i\omega^2k  \over
(1-ik\mu)\left\lbrack \alpha\,-\, {2k\over \sin
kL}\,(\cos\Phi-\cos kL)\, \right\rbrack -i\omega^2k}
   \end{equation}
for the reflection amplitude, in particular,
   \begin{equation} \label{delta reflection}
r(k)\,=\,-\,{(\alpha+ik)\sin kL-2k(\cos\Phi -\cos kL) \over
(\alpha-ik)\sin kL-2k(\cos\Phi -\cos kL)}
   \end{equation}
in the $\,\delta$--coupling case. This ($\,1\times 1\,$) S--matrix can
be also written by means of the phase shift. For instance, denoting
   \begin{equation} \label{delta denominator}
\Delta(k)\,\equiv\, \Delta(\alpha,\Phi;k)\,:=\, (\alpha-ik)\sin
kL-2k(\cos\Phi -\cos kL)\,,
   \end{equation}
we can write the \rhs of (\ref{delta reflection}) as
$\,e^{2i\delta(k)}\,$ with
   \begin{equation} \label{delta phase shift}
\delta(k)\,=\,{\pi\over 2}\,+\,\arctan {k\, \sin kL\over \re\Delta(k)}\:.
   \end{equation}
As usual the growth of the phase shift is related to the number of
scattering resonances within a given energy interval. It is clear
from (\ref{delta phase shift}) that $\,\delta(k)\,$ passes odd multiples of
$\,\pi/2\,$ whenever the denominator (\ref{delta denominator}) passes
zero, of course, when there is not a simultaneous zero in the
numerator. The last named situation occurs if and only if the flux
$\,\Phi\,$ through the loop is a multiple of $\,\pi\,$. Hence ``one
half" of resonances is missing in that case; similar conclusions can
be made in the general case of boundary conditions (\ref{reflection})
when
   \begin{equation} \label{phase shift}
\delta(k)\,=\,{\pi\over 2}\,+\,\arctan \left\lbrace\, \mu k+\,
{\omega^2 k\over \alpha\,-\, {2k\over \sin kL}\,(\cos\Phi-\cos kL)}\,
\right\rbrace\,.
   \end{equation}
This is related to the existence of embedded eigenvalues at
integer/halfinteger values of $\,\phi\,$ which will be clearly seen
from the resonance pole trajectories discussed below. The bound
states can also be found directly:
   \begin{description}
\item{\em (a)} It is clear that {\em positive--energy} bound states
may be supported only at the loop. If we restrict our attention to
the nontrivial case $\,\omega\ne 0\,$, this is possible in view of
(\ref{4 parameter bc}) when $\,u(0)= u'(0)\!-u'(L)=0\,$. Hence such
bound states exist only at integer/halfinteger values of the magnetic
flux (in the natural units) and the corresponding eigenfunctions are
given by (\ref{chi_n}) with an even $\,n\,$ for $\,\phi\,$ integer
and odd $\,n\,$ for $\,\phi\,$ halfinteger.
\item{\em (b)} In addition, there can be {\em negative} eigenvalues.
To find them we suppose that the loop wavefunction is given by the
first part of (\ref{Ansatz}) with $\,k=i\kappa\,$ and the halfline
part is $\,\rho\, e^{-\kappa x},\; \kappa>0\,$. The boundary
conditions then yield a system of equations for $\,\beta,\rho\,$
which can be solved provided
   \begin{equation} \label{negative bs}
{2\kappa\over \sinh\kappa L}\, (\cos\Phi- \cos\kappa L)\,=\,
\alpha+\, {\omega^2\kappa\over 1+\mu\kappa}\,.
   \end{equation}
It is easy to see that under the condition $\,\alpha\ge {2\over L}\,
(\cos\Phi-1)\,$ has no solution if $\,\mu\ge 0\,$ and a single root
otherwise; in the case $\,\alpha< {2\over L}\,(\cos\Phi-1)\,$ one
more eigenvalue is added.
   \end{description}

\section{A digression: a duality for graph scattering}

\setcounter{equation}{0}

At this point we want to make a small detour to describe a general
method to treat scattering problem on graphs. Recall that there is an
equivalence between the spectral problem for one--dimensional
Schr\"odinger operators with point interactions and certain Jacobi
matrices which is known in the literature as a ``French connection"
\cite{AGHH,BFLT,DSS,GH,GHK,Ph}. We have been able to extend this
duality recently to a wide class of Schr\"odinger operators on
graphs \cite{E5}; here we want to illustrate that the same method is
applicable to scattering problems.

\subsection{Schr\"odinger operators on a general graph}

Let us first collect some notion we shall need to formulate the
result; for more details we refer to \cite{E5}.  A graph $\,\Gamma\,$
consists of a finite or countably infinite number of {\em vertices}
$\,\VV=\{\XX_j:\,j\in I\}\,$  and {\em links (edges)}
$\,\LL=\{\LL_{jn}:\, (j,n)\in I_{\LL}\subset I\times I\}\,$.  Without
loss of generality we may suppose that each pair of vertices is
connected by not more than one link; otherwise we just add some
number of extra vertices. We assume that $\,\Gamma\,$ is connected,
so the set $\,\NN(\XX_j)= \{\XX_n:\, n\in\nu(j)\subset
I\setminus\{j\}\}\,$ of {\em neighbors} of $\,\XX_j\,$, \ie, the
vertices connected with $\,\XX_j\,$ by a single link, is nonempty.
Throughout we shall assume that $\,\NN(\XX_j)\,$ is {\em finite} for
any $\,j\in I\,$.

The graph {\em boundary} $\,\BB\,$ is the subset of vertices having a
single neighbor; it may be empty. We use the symbols $\,I_\BB\,$ and
$\,I_\II\,$ for the index subsets in $\,I\,$ corresponding to
$\,\BB\,$ and the graph {\em interior} $\,\II:=\VV\setminus\BB\,$,
respectively.  $\,\Gamma\,$ has a {\em local} metric structure coming
from the fact that each link $\,\LL_{jn}\,$ can be mapped to a line
segment $\,[0,\ell_{jn}]\,$. It is also possible to equip the graph
naturally with a {\em global} metric by identifying it with a subset
of a plane or a higher dimensional Euclidean space. The two metrics
may differ at a single link; the local one which is important for us
is usually given by the arc length of the curve segment representing
$\,\LL_{jn}\,$.

Using the local metric, we are able to introduce the state Hilbert
space in the way we did it for the lasso graph and similar problems,
namely as $\,L^2(\Gamma):= \bigoplus_{(j,n)\in I_\LL}
L^2(0,\ell_{jn})\,$. Its elements, \ie, the wave functions, will be
written as $\,\psi=\{\psi_{jn}:\, (j,n)\in I_\LL\}\,$
or simply as $\,\{\psi_{jn}\}\,$. We shall suppose that the particle
living on $\,\Gamma\,$ is exposed to a potential; it is only
important to know its values on the graph links, \ie, a family of
functions $\,V:=\{V_{jn}\}\,$; since we do not want deal with
mathematical subtleties here, we suppose that all of them are
{\em essentially bounded,} $\,V_{jn}\in L^{\infty}(0,\ell_{jn})\,$.
Then we are able to define the operator $\,H_\alpha\equiv
H(\Gamma,\alpha,V)\,$ by
   \begin{equation} \label{graph SO}
H_\alpha\{\psi_{jn}\} \,:=\, \{\, -\psi''_{jn}+V_{jn}\psi_{jn}\,:\;
(j,n)\in I_\LL\,\}
   \end{equation}
with the domain consisting of all $\,\psi\,$ with $\,\psi_{jn}\in
W^{2,2}(0,\ell_{jn})\,$ subject to a set $\,\alpha\,$ of boundary
conditions at the vertices which couple the boundary values
   \begin{equation} \label{boundary values}
\psi_{jn}(j):=\lim_{x\to 0+} \psi_{jn}(x)\,, \qquad
\psi'_{jn}(j):= \lim_{x\to 0+} \psi'_{jn}(x)\;;
   \end{equation}
we have identified here $\,x=0\,$ with the vertex $\,\XX_j\,$. In
general, there is vast family of boundary conditions which make the
operator (\ref{graph SO}) self--adjoint. It can be characterized by
$\,4M^2$ real parameters, where $\,M\,$ is the number of graph links
\cite{AGHH,ES,RS}, and even if we restrict to {\em local} boundary
conditions which do not couple the boundary values belonging to
different vertices, the number is still too large.

As above we restrict ourselves to the simplest situation when the
links connected in a vertex $\,\XX_j\,$ satisfy the
$\,\delta${\em--coupling} condition, \ie,
$\,\psi_{jn}(j)=\psi_{jm}(j)=:\psi_j\,$ for all $\,n,m\in\nu(j)\,$,
and
   \begin{equation} \label{delta N}
\sum_{n\in\nu(j)} \psi'_{jn}(j)\,=\, \alpha_j\psi_j
   \end{equation}
with a real--valued parameter $\,\alpha_j\in\R\;$ (coupling
constant). However, the results derived below can be reformulated
easily for the case when (\ref{delta N}) is replaced by a
$\,\delta'$--coupling or another type of local boundary conditions
\cite{E3,ES}.

As in the particular case discussed in the previous sections the
relation (\ref{delta N}) and other local couplings have an
illustrative meaning of probability current conservation at the
vertex; in a sense they represent an analogy of Kirchhoff's law.
This means, in particular, that they are independent of the lengths
of the involved links. Moreover, since the probability current is
connected with the kinetic part of the Schr\"odinger equation, the
coupling is also independent of the potential $\,V\,$ as long as the
latter is regular, which is the assumption we have adopted. At the
graph boundary we employ the usual conditions
   \begin{equation} \label{free end bc}
\psi_j \cos\omega_j+\psi'_j\sin\omega_j\,=\,0
   \end{equation}
with a parameter $\,\omega_j\,$; integer and halfinteger multiples
of $\,\pi\,$ correspond to the Dirichlet and Neumann condition,
respectively.

\subsection{Coupling two link bundles}

In the next step we attach a certain number of semiinfinite links to
$\,\Gamma\,$ which will support asymptotic solutions; in the standard
stationary picture we shall consider a combination of a falling and
transmitted/reflected plane wave on each of them. We might regard
these ``external" links as a part of the graph boundary; however, it
is convenient to treat them separately.  A reason for that is the
following: while we declared the intention to formulate the result
for graphs with $\,\delta$--couplings, it is desirable to have a
coupling between the internal and external links which is slightly
more general than (\ref{delta N}). This could be useful, \eg, if we
want to study perturbatively resonances which arise when eigenvalues
of the original graph operator become embedded into the continuous
spectrum of the leads.

As another preliminary, therefore, consider two bundles of
leads which support wavefunctions $\,\{f_n\}_{n=1}^N\,$ and
$\,\{g_m\}_{m=1}^M\,$; the endpoints are placed to the point
$\,x=0\,$. Suppose first that we have separate $\,\delta$--couplings
for each bundle,
   \begin{equation} \label{bundle continuity}
f_1(0)=\cdots=f_N(0)=:f(0)\,, \qquad g_1(0)=\cdots=g_M(0)=:g(0)\,,
   \end{equation}
together with $\,\sum_{n=1}^N f'_n(0)=\alpha f(0)\,$ and
$\,\sum_{m=1}^M g'_n(0)=\tilde\alpha g(0)\,$. To couple the
two bundles, we preserve the separate continuity (\ref{bundle
continuity}) and replace the derivative conditions by
   \begin{equation} \label{bundle flow}
f(0)\,=\, \alpha^{-1} \sum_{n=1}^N f'_n(0)+ \gamma \sum_{m=1}^M
g'_n(0)\,,  \;\quad g(0)\,=\, \bar\gamma \sum_{n=1}^N f'_n(0)+
\tilde\alpha^{-1}  \sum_{m=1}^M g'_n(0)
   \end{equation}
with a complex parameter $\,\gamma\,$; an elementary integration by
parts then shows that the corresponding boundary form vanishes under
these conditions. The parameter modulus is the coupling strength; if
the coupling is required to be time--reversal invariant, $\,\gamma\,$
has to be real. An overall $\,\delta$--coupling is achieved, of
course, if $\,\alpha=\tilde\alpha= \gamma^{-1}$.

\subsection{The S--matrix equation}

Suppose now that a bundle of $\,m_j\,$ halflines, $\,0\le m_j<
\infty\,$, is attached to the point $\,\XX_j\,$ of $\,\Gamma\,$; the
coupling being given by (\ref{bundle continuity}), (\ref{bundle
flow}) with the parameters $\,\alpha_j\,$ for the graph links joined at
$\,\XX_j\,$, $\;\tilde\alpha_j\,$ for the external links at
$\,\XX_j\,$, and $\,\gamma_j\,$. We call the $\,j$--th bundle
$\,\EE_j\,$ and $\,\EE_{jm}\,$ will be the $\,m$--th halfline in it,
so the full state Hilbert space will be now $\,L^2(\Gamma)\oplus
\left(\bigoplus_{j\in I} \bigoplus_{m=1}^{m_j} L^2(\EE_{jm}
\right)\,$. For the sake of brevity, the graph extended by the
external links will be denoted as $\,\Gamma_e\equiv \Gamma\cup\EE\,$,
for the state Hilbert space we will use the shorthand
$\,L^2(\Gamma_e)\,$. The symbol $\,H_\alpha\equiv
H(\Gamma_e,\{\alpha_j, \tilde\alpha_j,\gamma_j \},V)\,$ means a
Schr\"odinger operator on $\,\Gamma_e\,$ with the described coupling;
for simplicity we assume that the potentials on the external links
are zero.

As usual the stationary scattering problem means finding a
generalized eigenvector of $\,H_{\alpha}\,$ with prescribed behavior
in the asymptotic region, \ie, a solution to the equation
   \begin{equation} \label{local SO}
H_\alpha\psi\,=\, k^2\psi\,,
   \end{equation}
which belongs {\em locally} to $\,D(H_{\alpha})\,$ satisfying all the
domain requirement (in particular, the boundary conditions at each
vertex) apart of the global square integrability, and such that
   \begin{equation} \label{graph asymptotics}
\psi_{jm}(x)\,=\, a_{jm}e^{-ikx}+ b_{jm}e^{ikx}
   \end{equation}
holds for $\,x\in \EE_{jm}\,$. The vectors $\,a\equiv \{a_{jm}\}\,$
and $\,b\equiv \{b_{jm}\}\,$ of dimension $\,\card\EE= \sum_{j\in I}
m_j\,$ represent the incoming and outgoing amplitudes, respectively;
we are interested in the operator that maps the former into the
latter, $\,b=Sa\,$.

To proceed further, we need some more notation. The symbol
$\,H_\alpha^D\,$ will denote the decoupled operator obtained from
$\,H_\alpha\,$ by changing the conditions (\ref{delta N}) at the
points of graph interior $\,\II\,$ to Dirichlet, while at the
boundary they are kept fixed; we also define $\,\KK_\alpha:=\{ k:\,
k^2\in \sigma(H_\alpha^D),\: \im k\ge 0\,\}\,$. Next we take an
arbitrary link $\,\LL_{nj}\equiv [0,\ell_{jn}]\,$ of $\,\Gamma\,$,
the right endpoint being identified with $\,\XX_j\,$, and denote by
$\,u_{jn},\,v_{jn}\,$ the normalized Dirichlet solutions to the
corresponding component $\,-f''+V_{jn}f=k^2f\,$ of the Schr\"odinger
equation (\ref{local SO}). In other words, we demand that the
following boundary conditions are satisfied,
   \begin{equation} \label{normalized Dirichlet}
u_{jn}(\ell_{jn})= 1\!-\!u'_{jn}(\ell_{jn})=0\,,
\qquad v_{jn}(0)= 1\!-\!v'_{jn}(0)=0\,,
   \end{equation}
provided $\,n\in I_\II\,$; at the graph boundary we replace the last
requirement by $\,v_{jn}(0)=\sin\omega_n\,$ and
$\,v_{jn}'(0)=-\cos\omega_n \,$. The Wronskian of these solutions
equals
   \begin{equation} \label{Wronskian}
W_{jn}= -v_{jn}(\ell_{jn})= u_{jn}(0)
   \end{equation}
for $\,n\in I_\II\,$ and $\,W_{jn}=-u_{jn}(0) \cos\omega_n
-u'_{jn}(0)\sin\omega_n\,$ otherwise. All these quantities depend
in general on the spectral parameter $\,k\,$ but we shall not
indicate this fact explicitly. Now we can formulate the mentioned
result:
\vspace{3mm}

\noindent
{\bf Proposition.} {\em Let $\,k\not\in\KK_\alpha\,$ with
$\,k^2\in\R$, $\,\im k\ge 0\,$. Under the assumptions given above,
the corresponding on--shell scattering matrix for the graph
$\,\Gamma_e\,$ is given by the following system of $\,N:=\card
I+\card\EE\,$ equations
   \begin{eqnarray}
\sum_{n\in\nu(j)\cap I_\II} {\psi_n\over W_{jn}} \!&-&\!
\left(\, \sum_{n\in\nu(j)} {v'_{jn}(\ell_{jn})
\over W_{jn}} -\alpha_j\, \right)\psi_j -\, ik\alpha_j \gamma_j
m_j b_{j1} \nonumber \phantom{AAAAAA} \\ && \label{S duality a} \\
\!\!&=&\!\! -ik\alpha_j \gamma_j \left( m_j\alpha_{j1}- 2\sum_{m=2}^{m_j}
a_{jm} \right) \nonumber \\ \nonumber \\
\tilde\alpha_j \bar\gamma_j\Bigg( \sum_{n\in\nu(j)\cap I_\II}
{\psi_n\over W_{jn}}\!\!&-&\!\! \sum_{n\in\nu(j)} {v'_{jn}(\ell_{jn})
\over W_{jn}} \,\Bigg)\psi_j +\, b_{j1}\left( \tilde\alpha_j
-ikm_j \right) \nonumber \\ && \label{S duality b} \\
\!&=&\! -a_{j1} \left( \tilde\alpha_j +ikm_j \right)
- 2ik \sum_{m=2}^{m_j} a_{jm} \nonumber
   \end{eqnarray}
and}
   \begin{equation} \label{S continuity}
b_{jm} =b_{j1}+a_{j1}-a_{jm}\;, \qquad m=2,\dots, m_j\,.
   \end{equation}
\vspace{2mm}

\noindent
{\bf Remarks.} (a) If $\,N<\infty\,$ the above relations represent a
system of linear equations. In the opposite case they have to be
interpreted as the appropriate operator equation on $\,\ell^2$. This
can be done under some additional assumptions on $\,\Gamma\,$, \eg,
if there are positive numbers $\,c_1,c_2\,$ such that $\,c_1\le
\ell_{jn}\le c_2\,$ holds for all $\,(j,n)\in I_\LL\,$ --- see
\cite{E5} for more details.

(b) The results generalizes easily to the situation when $\,\Gamma\,$
as a subset of $\,\R^\nu$ is placed into a magnetic field, not
necessarily homogeneous, described by a vector potential $\,A\,$. The
boundary conditions (\ref{delta N}) are modified replacing
$\,\psi'_{jn}(j)\,$ by $\,\psi'_{jn}(j) +iA_{jn}(j)\,$, where
$\,A_{jn}(j)\,$ is the tangent component of $\,A\,$ to $\,\LL_{jn}\,$
at $\,\XX_j\;$ \cite{ARZ}. The particle abiding on $\,\Gamma\,$ is
supposed here to be an electron; otherwise $\,A\,$ has to be replaced
by $\,-qA\,$ where $\,q\,$ is the particle charge. The magnetic case
can be handled by means of the unitary operator $\,U:\:
L^2(\Gamma)\to L^2(\Gamma)\,$ which acts as
$$
(U\psi)_{jn}(x)\,:=\, \exp\left( i\int_{x_{jn}}^x A_{jn}(y)\,dy
\right)\, \psi_{jn}(x)\;;
$$
the values $\,x_{jn}\,$ are fixed reference points. Then the
functions $\,(U\psi)_{jn}\,$ satisfy (\ref{delta N}) and
it is sufficient to replace the function values $\,\psi_n\,$ in
(\ref{S duality a}), (\ref{S duality b}) by $\,e^{iA_n} \psi_n\,$
provided the magnetic phase factors $\,A_j\,$ are chosen to obey the
natural consistency condition
$$
A_j\!-A_n\,=\, \int_{\LL_{jn}} A_{jn}(y)\, dy\,.
$$
required by the wave function continuity.

(c) Consider a simple situation when a single halfline is attached to
every point of $\,\Gamma\,$ and denote the ``graph part" of the above
system, \ie, the operator represented by the two sums at the \lhs of
(\ref{S duality a}) as $\,h\,$. If the coupling is ideal,
$\,\alpha_j=0\,$ for all $\,j\in I\,$, the S--matrix is given by
$$
S\,=\,-\, {h+ik\over h-ik}\,.
$$
It is illustrative to compare this to the the formula used recently
by Sadun and Avron \cite{SA} in a study of scattering on discrete
graphs; the only difference is the replacement of $\,-ik\,$ by
$\,e^{ik}$, the energy being $\,2\cos k\,$ in this case.
\vspace{3mm}

To prove the proposition, it is sufficient to use the transfer
matrices which relate the Schr\"odinger equation solutions at both
ends of each link \cite{E5}. Since the Wronskian is nonzero for
$\,k\not\in\KK_\alpha\,$, we get
   \begin{eqnarray*}
\psi_j &\!\!:=\!\!& \psi_{jn}(j)= u'_{jn}(0)\psi_n+
v_{jn}(\ell_{jn})\psi'_{jn}(n)\,, \\ \\
-\psi'_{jn}(j) &\!=\!& {{1\!-\!u'_{jn}(0)v'_{jn}(\ell_{jn})}
\over W_{jn}} \psi_n+ v'_{jn}(\ell_{jn})\psi'_{jn}(n)\;;
   \end{eqnarray*}
the sign change at the \lhs of the last condition reflects the fact
that (\ref{boundary values}) defines the {\em outward} derivative at
$\,\XX_j\,$. We express $\,\psi'_{jn}(n)\,$ from the first relation
and substitute to the second one. This yields
$$
\psi'_{jn}(j)\,=\, -\,{\psi_n\over W_{jn}}\,+\, {v'_{jn}(\ell_{jn})
\over W_{jn}}\, \psi_j
$$
for $\,n\in I_\II\,$, while at the graph boundary we get with the
help of (\ref{free end bc}) instead
$$
\psi'_{jn}(j)\,=\,{v'_{jn}(\ell_{jn}) \over W_{jn}}\, \psi_j\,.
$$
Now one has just to substitute these values into the boundary
conditions at each vertex to arrive at the relations (\ref{S duality
a})--(\ref{S continuity}). \quad \QED
\vspace{3mm}

It is not difficult to check that the lasso graph with the
$\,\delta$--coupling can be treated within this general scheme. We
use the normalized Dirichlet solutions at both loop ``ends",
$\,k^{-1}\sin kx\,$ and $\,-k^{-1} \sin k(x\!-\!L)\,$, and add a
vertex into an ``interior point". Using (\ref{S duality a}) and
(\ref{S duality b}) and excluding the function values at the added
point, we arrive after a straightforward calculation at the equations
\begin{eqnarray*}
\left( -k\, {e^{i\Phi}+ e^{-i\Phi}\over \sin kL}\,+\, 2k\, {\cos
kL\over \sin kL}\,+\, \alpha \right)\psi -ikb \!&=&\!
-ika\,, \\ \\
\psi \!&=&\! b+a\,,
   \end{eqnarray*}
from which we recover the reflection amplitude (\ref{delta
reflection}).

\section{Resonances}

\setcounter{equation}{0}

\subsection{The resolvent}

Let us return now to our model. The most natural way to study
spectral properties of an operator is through its resolvent, and
therefore we want to find it for $\,H_{\alpha}(B)\,$. The
``decoupled" resolvent is found easily: it is a matrix integral
operator with the kernel
   \begin{equation} \label{decoupled kernel}
G_{\infty}(x,y;k)\,=\, \left( \begin{array}{cc} e^{-iA(x-y)}\, {\sin
kx_< \sin k(x_>-L) \over k\,\sin kL} & 0 \\ \\
0 & {\sin kx_<\, \exp(ikx_>) \over k} \end{array} \right)\;,
   \end{equation}
where $\,x_<\,$ and $\,x_>\,$ mean conventionally the smaller and
larger of the variables $\,x,\,y\,$, respectively. We abuse here
again the notation and employ the same symbol for the arc--length
variable on the loop and the lead as well as for the pair of them.

Since $\,H_{\alpha,\mu,\omega}\,$ and $\,H_{\infty}\,$ are both
self--adjoint extensions of the same symmetric operator with the
deficiency indices $\,(2,2)\,$, the resolvent of the former is by
Krein's formula \cite[App.A]{AGHH} given by
   \begin{equation} \label{Krein}
G_{\alpha,\mu,\omega}(x,y;k)\,=\, G_{\infty}(x,y;k)+
\sum_{j,\ell=1}^2 \lambda_{j\ell} F_j(x)F_{\ell}^t(y)\,,
   \end{equation}
where the symbol {\em ``t"} means transposition, $\,F_j\,$ are
vectors of the corresponding deficiency subspaces which we shall
choose in the form
   \begin{equation} \label{deficiency vectors}
F_1(x):= {w(x) \choose 0}\,, \qquad F_2(x):= { 0 \choose e^{ikx} }
   \end{equation}
with
$$
w(x) := e^{iAx}\, {e^{-i\Phi} \sin kx -\sin k(x\!-\!L) \over \sin
kL}\,,
$$
and $\,\lambda_{j\ell}\,$ are coefficients to be found. Introducing
$$
h_1:= \int_0^L w(y)v(y)\, dy\,, \qquad h_2:= \int_0^{\infty}
e^{iky}g(y)\, dy
$$
for a given $\,{v\choose g}\in\HH\,$, we find easily that the
boundary values of the function $\,{u\choose f}:= \left(
H_{\alpha,\mu,\omega}\!-k^2 \right)^{-1} {v\choose g}\,$ are in view of
(\ref{Krein}) given by
\begin{eqnarray*}
u(0) \!&=&\! u(L) = \lambda_{11}h_1\!+\lambda_{12}h_2\,, \qquad f(0)=
\lambda_{21}h_1\!+\lambda_{22}h_2\,, \\ \\
u'(0)-u'(L) \!&=&\! h_1+ {2k\over \sin kL}\, (\cos\Phi- \cos kL)
\left( \lambda_{11}h_1\!+\lambda_{12}h_2 \right)\,, \\ \\
f'(0)\!&=&\! h_2+ ik\left( \lambda_{21}h_1\!+\lambda_{22}h_2\right)\,.
\end{eqnarray*}
However, $\,{u\choose f}\,$ belongs to
$\,D(H_{\alpha,\mu,\omega})\,$ for any $\,{v\choose g}\,$, so
substituting these boundary values into (\ref{4 parameter bc}) we get
a system of four linear equations which yields the sought
coefficients:
   \begin{eqnarray}
\lambda_{11} \!&=&\! -\, {1-i\mu k\over \DD}\,, \qquad \lambda_{12}=
-\, {\omega\over\DD}\,, \nonumber \\ \label{coefficients} \\
\lambda_{21}\!&=&\! -\, {\omega\over\DD}\,, \qquad \lambda_{22} =
{\mu \left\lbrack 2k\, {cos\Phi-\cos kL\over \sin kL}\,-\alpha
\right\rbrack - \omega^2 \over \DD} \nonumber
   \end{eqnarray}
with
   \begin{equation} \label{discriminant}
\DD\equiv \DD(\alpha,\mu\,\omega;k)\,:=\, (1-i\mu k) \left\lbrack
2k\, {cos\Phi-\cos kL\over \sin kL}\,-\alpha \right\rbrack
-i\omega^2 k\,.
   \end{equation}
In the case of $\,\delta$--coupling, $\,\mu=0,\; \omega=1,\,$ the
coefficients acquire a particularly simple form, $\,\lambda_{j\ell}=
-\DD^{-1}, \; j,\ell=1,2\,$.

\subsection{Pole trajectories}

As usual in such situations \cite{AESS,ESe} the singularities of
$\,G_{\infty}(x,y;k)\,$ cancel with those of the added term in
(\ref{Krein}) and the resolvent poles are given by zeros of the
denominator (\ref{discriminant}); the exception is represented by the
case of an integer or halfinteger $\,\phi\,$.

For the sake of simplicity, we shall speak mostly about the
$\,\delta$--coupling situation. If the coupling is ideal,
$\,\alpha=0\,$, the pole condition becomes
   \begin{equation} \label{alpha zero poles}
2(\cos\Phi- \cos kL)= -i\,\sin kL
   \end{equation}
and one is able to solve it explicitly. No singularities exist in the
upper halfplane, hence we write
   \begin{equation} \label{pole position}
k\,=\, \kappa-i\eta\,.
   \end{equation}
Substituting into the above condition, we find that for $\,|\Phi|<
{\pi\over 6}\; ({\rm mod\;} \pi)\,$ there is a pair of poles with
$\,\kappa= \pi n/L\,$ and
   \begin{equation} \label{alpha zero}
\eta= {1\over L}\, \ln\left(2(-1)^n\cos\Phi\pm
\sqrt{4\cos^2\Phi-3} \right)\,,
   \end{equation}
where $\,(-1)^n\cos\Phi>0\,$. On the other hand, for the remaining
values of $\,\Phi\,$ the poles are found at the line parallel to the
real axis with $\,\eta=\, -\, {\ln 3\over 2L}\,$ and
   \begin{equation} \label{alpha zero'}
\kappa=\,\pm\, {1\over L}\, \arccos\left({2\over\sqrt 3}\, \cos\Phi
\right)\,.
   \end{equation}
We see that both poles are in the open lower halfplane with the
exception of $\,\Phi=n\pi\,$, \ie, $\,\phi\,$ integer or
halfinteger, when one of them turns into an embedded--eigenvalue pole
at the real axis. The pole trajectories with respect to $\,\Phi\,$
are not smooth despite the analytic form of the condition (\ref{alpha
zero poles}); this is due to the fact that $\,\DD=0\,$ at the crossing
points $\,{1 \over L}\left(\pi n-\,{i\over 2}\ln 3\right)\,$, so the
implicit--function theorem does not apply there. A similar picture is
obtained for the boundary conditions (\ref{4 parameter bc}) with
$\,\alpha=\mu=0\,$ and $\,|\omega|<\sqrt 2\,$, in which case the
``horizontal" line has $\,\eta= {1\over 2}\, \ln{2+\omega^2\over
2-\omega^2}\,$. On the other hand, in the case $\,|\omega|\ge \sqrt
2\,$ the pole trajectories are ``vertical" segments with $\,\kappa=
\pi n/L\,$ only.

If $\,\alpha\ne 0\,$ the $\,\delta$--coupling pole condition
(\ref{alpha zero poles}) is replaced by
$$
2k(\cos\Phi- \cos kL)= (\alpha-ik)\sin kL\,.
$$
Writing separately the real and imaginary parts with the help of the
parametrization (\ref{pole position}), we find that for $\,\kappa=\pi
n/L\,$ a zero can exist only at the real axis if $\,\Phi=m\pi\,$. For
other values of $\,\kappa\,$ the pole condition can be cast into the
form
   \begin{equation} \label {alpha nonzero}
\coth\eta L= 2+\, \alpha\, {2\eta-\kappa\cot\kappa L\over
\eta(\eta\!-\! \alpha)+\kappa^2}\,,
   \end{equation}
which has to be solved numerically. The resulting pole trajectories
are shown on Figure~2.

   \begin{figure}
   % INSERT FIGURE 2
   \caption{Pole trajectories from the condition (5.10) for different
values of the coupling constant (dashed: $\,\alpha=0.5\,$, full:
$\,\alpha=0.1\,$, dotted: $\,\alpha=0.05\,$) }
   \end{figure}

\section{Decay of loop states}

\setcounter{equation}{0}

Up to now we have considered the lasso graph as a scattering system.
Now we shall suppose that the system is prepared at an initial
instant in a state the wavefunction of which is localized at the
loop. It is not so important how such a situation is realized. For
instance, one can place an electron at an isolated ring and ``switch
in" the junction at $\,t=0\,$. The state is generally unstable under
the evolution governed by $\,H_{\alpha,\mu,\omega}$ and we are
interested in the way in which it decays.

Since we have an explicit expression for the resolvent, we are able
in principle to write the non--decay amplitude explicitly
\cite[Sec.3.1]{E1}. However, instead of trying to evaluate this function
we limit ourselves to elucidation of its basic properties.

\subsection{Spectral decomposition}

The relations (\ref{Krein}) and (\ref{coefficients}) imply, in
particular, that the resolvent form $\,z\mapsto \left(\psi, \left(
H_{\alpha,\mu,\omega}\!-z \right)^{-1}\psi\right)\,$ is a meromorphic
function including its continuation to the second sheet. Its possible
poles are associated with the discrete spectrum of
$\,H_{\alpha,\mu,\omega}$ which we also know explicitly. Since these
are the only singularities, the function $\,\left(\psi, \left(
H_{\alpha,\mu,\omega}\!-\cdot \right)^{-1}\psi\right)\,$ is analytic
for all $\,\psi\,$ belonging to the complement
$\,\HH_p(H_{\alpha,\mu,\omega})^{\perp}$. In particular, it is
uniformly bounded in any finite part of the strip $\,|\im z|<1\,$,
and thus by the basic criterion \cite[Thm.XIII.19]{RS} such a vector
belongs to $\,\HH_{ac}(H_{\alpha,\mu,\omega})\,$.

Consequently, our Hamiltonian has no singularly continuous spectrum.
The initial state can be therefore decomposed into $\,\psi= \psi_p+
\psi_{ac}\,$ and the corresponding non--decay amplitude equals
   \begin{equation} \label{nondecay amplitude}
(\psi,U_t\psi)= (\psi_p,U_t\psi_p)+ (\psi_{ac},U_t\psi_{ac})\,,
   \end{equation}
where $\,U_t:= \exp\{-iH_{\alpha,\mu,\omega}t\}\,$.  The second term
on the \rhs goes to zero as $\,t\to\infty\,$ in view of the
Riemann--Lebesgue lemma; the first one is a linear combination of
exponentials with coefficients coming from the Fourier decomposition
of $\,\psi_p\,$. If just one of them is nonzero, then the decay law
of the corresponding loop state given by squared modulus of
(\ref{nondecay amplitude}) has a finite nonzero limit. Such a
behavior is typical for unstable systems having bound states with a
nonzero Fourier component in a decaying state; it has been observed
recently in another context --- see \cite{E6}, and also \cite{GS}
where, however, the effect may be also related to the threshold
violation of the Fermi golden rule discovered by Howland \cite{Ho}.

If the loop state contains a superposition of a larger number of
eigenvectors, the nondecay probability does not go to zero as
$\,t\to\infty\,$ but a limit does not exist. In view of the above
discussion, such a situation can occur in the present model only if
(a) there are two negative eigenvalues (see Remark~(b) at the end of
Section~3), or (b) if $\,\Phi=n\pi\,$ with $\,n\in\Z\,$. The
asymptotic behavior of the decay law depends then on the coupling
parameters. If all the involved eigenvalues are commensurate, the
asymptotics is periodic; this happens always if there is no
negative--energy bound state. In the general case the decay law
asymptotics is quasiperiodic.

\subsection{What has all this in common with neutral kaons?}

Concluding this study, let me mention one more topic to which Larry
Horwitz made a contribution, namely the decay theory of neutral
kaons. This subject attracted attention at the end of the sixties as
an example of a system with a substantially nonexponential decay law
exhibiting different time scales, as well as the possibility to
``recreate" decayed particles by performing a set of noncompatible
measurements.

Mesoscopic physics makes it possible to tailor systems in which
similar effect can be observed. Consider our lasso graph with the
initial wavefunction $\,u\,$ on the loop such that $\,x\mapsto
e^{iAx}u(x)\,$ has no definite symmetry with respect to the
connection point $\,x=0\;$ (say, $\,u(x)= e^{-ix(A-2\pi n/L)}\,$). If
the flux value $\,\phi\,$ is integer, the $\,A$--even component
represents a superposition of embedded--eigenvalue bound states and
thus it survives, while the $\,A$--odd one dies out. In a real life
experiment, of course, we cannot ensure that $\,\phi\,$ is exactly an
integer, hence we shall have rather a fast and a slowly decaying part
of the wavefunction; recall the pole trajectories discussed in
Section~5.2.

Moreover, consider a loop to which two halfline leads are attached at
different points and assume that we are able to switch the coupling
in and out independently. We wait until the $\,A$--odd part in the
above described experiment essentially decays while the longliving
component is still preserved, and switch from the first lead to the
second one. Now the symmetry with respect to the other junction is
important. If the surviving part of the wavefunction is a
superposition of an $\,A$--even and an $\,A$--odd part with respect
to the latter, the scenario repeats. Of course, the ``second decay"
may produce a smaller component $\,A$--odd with respect to the first
junction, so the analogy is complete.

\subsection*{Acknowledgment}

The trajectories featured on Figure~2 have been computed by M. Tater.
A partial support by the grants AS No.148409 and GACR
No.202--96--0218 is gratefully acknowledged. 

\subsection*{Figure captions}

   \begin{description}
   \item{\bf Figure 1.} A lasso graph
   \item{\bf Figure 2.} Pole trajectories from the condition (5.10)
for different values of the coupling constant (dashed:
$\,\alpha=0.5\,$, full: $\,\alpha=0.1\,$, dotted: $\,\alpha=0.05\,$)
   \end{description}


\begin{thebibliography}{article}
   \bibitem{Ad}
V.M.~Adamyan: Scattering matrices for microschemes, {\em Oper.Theory:
Adv. Appl.} {\bf 59} (1992), 1--10.
   \vspace{-1.8ex}
   \bibitem{AGHH}
S.~Albeverio, F.~Gesztesy, R.~H\o egh-Krohn, H.~Holden: {\em Solvable Models
in Quantum Mechanics}, Springer, Heidelberg 1988.
   \vspace{-1.8ex}
   \bibitem{AESS}
J.--P.~Antoine, P.~Exner, P.~\v Seba, J.~Shabani: A mathematical model
of heavy--quarkonia decays, {\em Ann.Phys.} {\bf 233} (1994), 1--16.
   \vspace{-1.8ex}
   \bibitem{AL}
Y.~Avishai, J.M.~Luck: Quantum percolation and ballistic conductance
on a lattice of wires, {\em Phys.Rev.} {\bf B45} (1992), 1074--1095.
   \vspace{-1.8ex}
   \bibitem{AEL}
J.E.~Avron, P.~Exner, Y.~Last: Periodic Schr\"odinger operators with
large gaps and Wannier--Stark ladders, {\em Phys.Rev.Lett.} {\bf 72}
(1994), 896--899.
   \vspace{-1.8ex}
   \bibitem{ARZ}
J.E.~Avron, A.~Raveh, B.~Zur: Adiabatic transport in multiply
connected systems, {\em Rev.Mod.Phys.} {\bf 60} (1988), 873--915.
   \vspace{-1.8ex}
   \bibitem{BFLT}
J.~Bellissard, A.~Formoso, R.~Lima, D.~Testard: Quasi--periodic
interactions with a metal--insulator transition, {\em Phys.Rev.}
{\bf B26} (1982), 3024--3030.
   \vspace{-1.8ex}
   \bibitem{Bu}
M.~B\"uttiker: Small normal--metal loop coupled to an electron
reservoir, {\em Phys.Rev.} {\bf B32} (1985), 1846--1849.
   \vspace{-1.8ex}
   \bibitem{DSS}
F.~Delyon, B.~Simon, B.~Souillard: From power pure point to
continuous spectrum in disordered systems, {\em Ann.Inst.
H.~Poincar\'e} {\bf A42} (1985), 283--309.
   \vspace{-1.8ex}
   \bibitem{E1}
P.~Exner: {\em Open Quantum Systems and Feynman Integrals}, D.
Reidel, Dordrecht 1984.
   \vspace{-1.8ex}
   \bibitem{E2}
P.~Exner: Lattice Kronig--Penney models, {\em Phys.Rev.Lett.} {\bf
74} (1995), 3503--3506.
   \vspace{-1.8ex}
   \bibitem{E3}
P.~Exner: Contact interactions on graph superlattices, {\em J.Phys.}
{\bf A29} (1996), 87--102.
   \vspace{-1.8ex}
   \bibitem{E4}
P.~Exner: Weakly coupled states on branching graphs, {\em
Lett.Math.Phys.}, to appear
   \vspace{-1.8ex}
   \bibitem{E5}
P.~Exner: A duality between Schr\"odinger operators on graphs and
certain Jacobi matrices, {\em Ann.Inst. H. Poincar{\'e}} {\bf 63}
(1996), to appear
   \vspace{-1.8ex}
   \bibitem{E6}
P.~Exner: A solvable model of two--channel scattering, {\em
Helv.Phys.Acta} {\bf 64} (1991), 592--609.
   \vspace{-1.8ex}
   \bibitem{EG}
P.~Exner, R.~Gawlista: Band spectra of rectangular graph
superlattices, {\em Phys.Rev.} {\bf B53} (1996), 7275--7286.
   \vspace{-1.8ex}
   \bibitem{ES}
P.~Exner, P.~\v{S}eba: Free quantum motion on a branching graph, {\em
Rep. Math.Phys.} {\bf 28} (1989), 7--26.
   \vspace{-1.8ex}
   \bibitem{ESe}
P.~Exner, E.~\v{S}ere\v{s}ov\'{a}: Appendix resonances on a simple
graph, {\em J. Phys.} {\bf A27} (1994), 8269--8278.
   \vspace{-1.8ex}
   \bibitem{GS}
B.~Gaveau, L.S.~Schulman: Limited quantum decay, {\em J. Phys.}
{\bf A28} (1995), 7359--7374.
   \vspace{-1.8ex}
   \bibitem{GP}
N.I.~Gerasimenko, B.S.~Pavlov: Scattering problem on noncompact graphs,
{\em Teor.Mat.Fiz.} {\bf 74} (1988), 345-359 (in Russian).
   \vspace{-1.8ex}
   \bibitem{GH}
F.~Gesztesy, H.~Holden: A new class of solvable models in quantum
mechanics describing point interactions on the line, {\em J.Phys.}
{\bf A20} (1987), 5157--5177.
   \vspace{-4.8ex}
   \bibitem{GHK}
F.~Gesztesy, H.~Holden, W.~Kirsch: On energy gaps in a new type of
analytically solvable model in quantum mechanics, {\em
J.Math.Anal.Appl.} {\bf 134} (1988), 9--29.
   \vspace{-1.8ex}
   \bibitem{GLRT}
J.~Gratus, C.J.~Lambert, S.J.~Robinson, R.W.~Tucker: Quantum
mechanics on graphs, {\em J.Phys.} {\bf A27} (1994), 6881--6892.
   \vspace{-1.8ex}
   \bibitem{HLM}
L.P.~Horwitz, J.A.~LaVita, J.--P.~Marchand: The inverse decay
problem, {\em J.Math.Phys.} {\bf 12} (1971), 2537--2543.
   \vspace{-1.8ex}
   \bibitem{HM1}
L.P.~Horwitz, J.--P.~Marchand: The decay scattering system, {\em
Rocky Mts.J.Math.} {\bf 1} (1971), 225--253.
   \vspace{-1.8ex}
   \bibitem{HM2}
L.P.~Horwitz, J.--P.~Marchand: Unitary sum rule and the time
evolution of neutral K--mesons, {\em Helv.Phys. Acta} {\bf 42}
(1969), 801--807.
   \vspace{-1.8ex}
   \bibitem{HM3}
L.P.~Horwitz, J.--P.~Marchand: Formal scattering treatment of the
neutral K meson system, {\em Helv.Phys. Acta} {\bf 42} (1969),
1039--1054.
   \vspace{-1.8ex}
   \bibitem{HS}
L.P.~Horwitz, I.M.~Sigal: On a mathematical model for non--stationary
physical system, {\em Helv. Phys. Acta} {\bf 51} (1978), 685--715.
   \vspace{-1.8ex}
   \bibitem{Ho}
J.S.~Howland: Puiseux series for resonances at embedded eigenvalues,
{\em Pacific J.Math.} {\bf 55} (1974), 157--176. 
   \vspace{-1.8ex}
   \bibitem{JS}
A.M.~Jaynnavar, P.~Singha Deo: Persistent current and conductance of
metal loop connected to electron reservoir, {\em Phys.Rev.} {\bf B49}
(1994), 13685--13690.
   \vspace{-1.8ex}
   \bibitem{Ph}
P.~Phariseau: The energy spectrum of an amorphous substance, {\em Physica}
{\bf 26} (1960), 1185--1191.
   \vspace{-1.8ex}
   \bibitem{RS}
M.~Reed, B.~Simon: {\em Methods of Modern Mathematical Physics,
I.~Functional Analysis, IV.~Analysis of Operators}, Academic Press,
New York 1972, 1978.
   \vspace{-1.8ex}
   \bibitem{RuS}
K.~Ruedenberg, C.W.~Scherr: Free--electron network model for
conjugated systems, I.~Theory, {\em J.Chem.Phys.} {\bf 21} (1953),
1565--1581.
   \vspace{-1.8ex}
   \bibitem{SA}
L.~Sadun, J.E.~Avron: Adiabatic curvature and the S--matrix,
preprint mp$\underline{\phantom{x}}$arc 95--518; to appear in {\em
Commun.Math.Phys.}. 
   \vspace{-1.8ex}
   \bibitem{Si}
B.~Simon: Almost periodic Schr\"odinger operators: a review,
{\em Adv.Appl. Math.} {\bf 3} (1982), 463--490.
   \vspace{-1.8ex}
   \end{thebibliography}
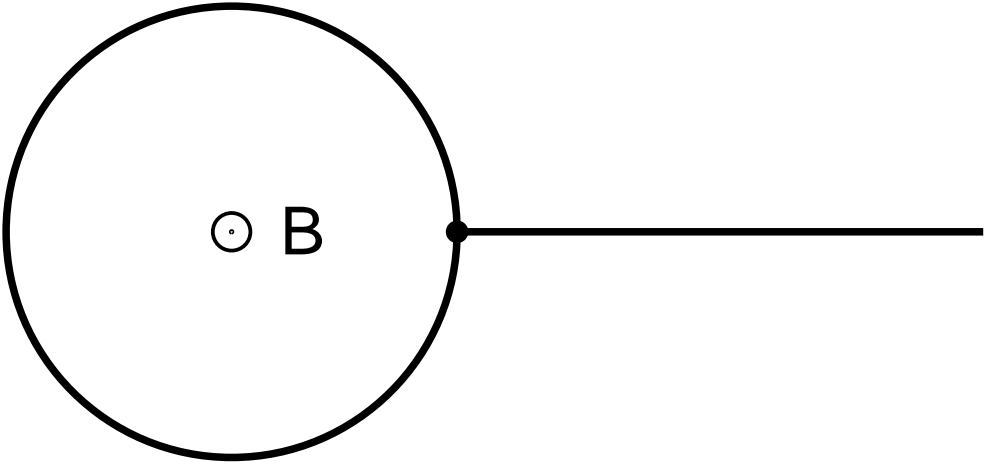
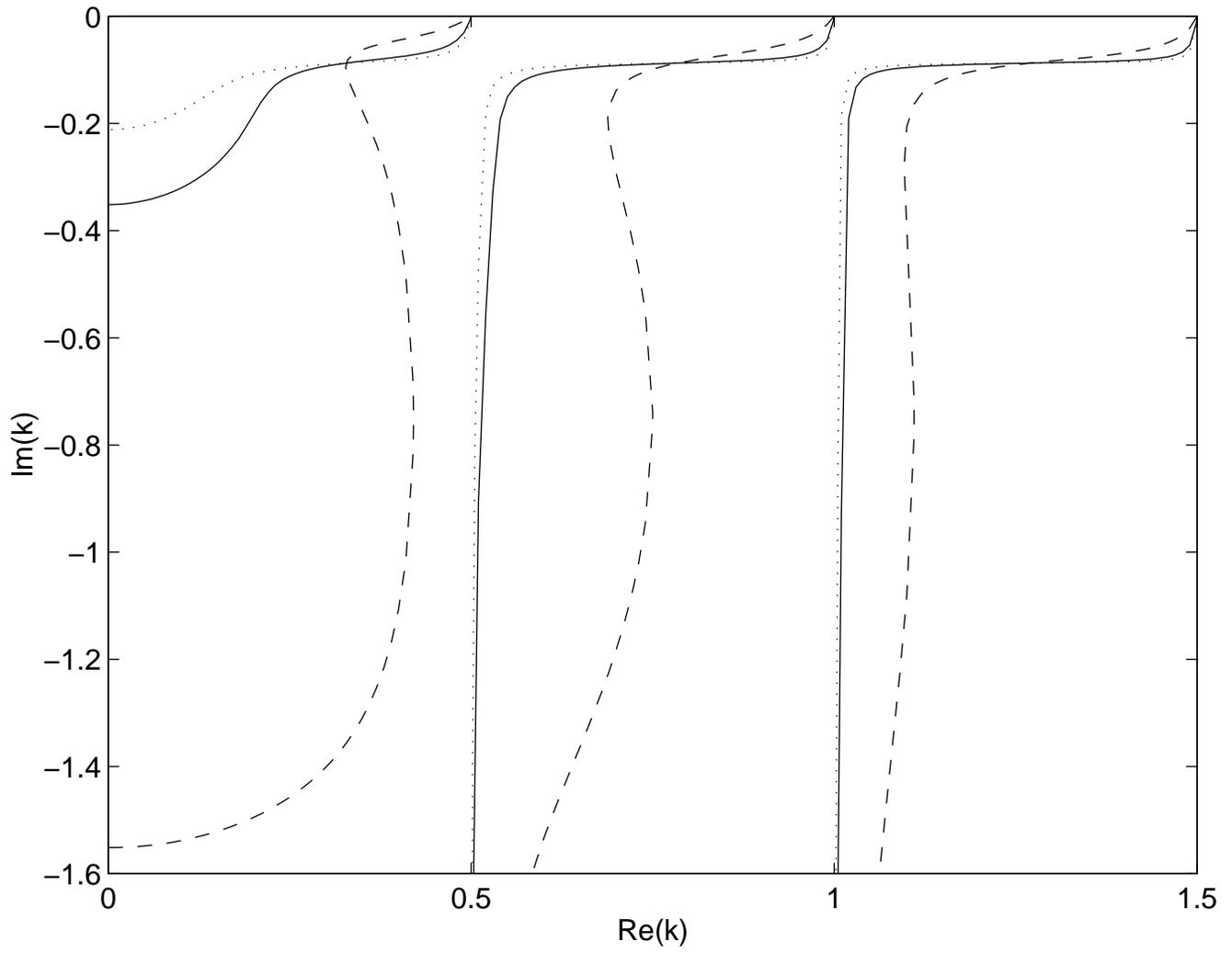
\end{document}